\documentclass[manuscript]{aastex}

\begin{document}
\pagenumbering{arabic}

\title{GLOBULAR CLUSTERS IN DWARF GALAXIES}

\author{Sidney van den Bergh}
\affil{Dominion Astrophysical Observatory, Herzberg Institute of Astrophysics, National Research Council of Canada, 5071 West Saanich Road, Victoria, BC, V9E 2E7, Canada}
\email{sidney.vandenbergh@nrc.gc.ca}

\begin{abstract}
 Data are presently available on the luminosities and 
half-light radii of 101 globular clusters associated with low- 
luminosity parent galaxies. The luminosity distribution of 
globulars embedded in dwarf galaxies having $M_{v} > -16$ is found to 
differ dramatically from that for globular clusters surrounding 
giant host galaxies with  $M_{v} < -16$. The luminosity distribution of 
globular clusters in giant galaxies peaks at $M_{v} \sim -7.5$, whereas 
that for dwarfs is found to increases monotonically down to the 
completeness limit of the cluster data at $M_{v} \sim -5.0$. Unexpectedly, 
the power law distribution of the luminosities of globular 
clusters hosted by dwarf galaxies is seen to be much flatter 
than the that of bright unevolved part of the luminosity 
distribution of globular clusters associated with giant galaxies. 
The specific frequency of globular clusters that are fainter than 
$M_{v} = -7.5$ is found to be particularly high in dwarf galaxies. The 
luminosity distribution of the LMC globular clusters is similar 
to that in giant galaxies, and differs from those of the globulars 
in dwarf galaxies. The present data appear to show no strong 
dependence of globular cluster luminosity on the morphological
types of their parent galaxies. No attempt is made to explain the
unexpected discovery that the luminosity distribution of globular
clusters is critically dependent on parent galaxy luminosity (mass?),
but  insensitive to the morphological type of their host galaxy.

 
\end{abstract}


\section{INTRODUCTION}

 It is the purpose of the present paper to see if the 
luminosity distribution of globular clusters depends on the 
luminosity (mass?) of their host galaxy. There are at least 
two reasons why such differences might be expected: (1) The 
evolutionary history of giant galaxies (which is believed to 
have been dominated by hierarchical merging) may have differed 
significantly from that of dwarf galaxies, and (2) cluster 
destruction by bulge and disk shocks is expected to be much 
more important in massive giants than it is in low mass dwarfs. 
Finally it is of interest to enquire if the luminosity 
distribution of the globular clusters hosted by dwarf 
galaxies is also a function of the morphological type of
their parent galaxy. Recent investigations by Sharina, Puzia \& Marakov (2005), by van den Bergh \& Mackey (2004) and by Mackey \& van den Bergh 
(2005) provide information on the luminosities $M_{v}$ and half-light 
radii $R_{h}$ of 101 globular clusters that are associated with host 
galaxies which are of intermediate or low luminosities. These 
data, which are collected in Table 1, have been restricted to 
hosts  with luminosities fainter than $M_{v} = -18.9$. This 
limit was chosen so as to include  the Large Magellanic Cloud 
($M_{v} = -18.5$), while excluding M 33 ($M_{v} = -18.9$), for which the 
observational data on globular clusters are still both inhomogeneous 
and incomplete. Table lists (1) the name of the parent galaxy, 
(2) the de Vaucouleurs type, (3) the luminosity of this host galaxy, 
(4) the name of each  cluster, (5) the absolute cluster magnitude $M_{v}$ 
and (6) the projected distance of each cluster from the center of its 
host galaxy. Uncertain values are followed by a colon (:). Angular 
dimensions were converted to pc or kpc using the distances 
listed in van den Bergh (2000). Clusters with intrinsic colors 
$(V-I)_{o} < 0.70$ were excluded because of the high probability that 
they might actually be young open clusters. In Table 1 host 
galaxies with distance moduli $(m-M)_{o} < 28.0$,  corresponding to 
a distance D $<$ 4.0 Mpc, have been marked with an asterisk (*).
The data for the more remote  galaxies with D $>$ 4.0 Mpc are 
probably more uncertain than are those for the globulars in nearer 
galaxies.

\section{DATA SAMPLE}

 Sharina et al. have used HST images of 57 nearby low
surface brightness galaxies to search for globular cluster
candidates. The majority of the  fields they searched were  
located in small clusterings of galaxies  (M81 group, Sculptor 
group, CVn I cloud,  Cen A group, NGC 3115 group,  while
35\% of their target galaxies were situated in the field. To
enlarge the data sample information  on globular clusters 
that are associated with dwarf members of  the Local Group  
which is, in many respects, similar to the small galaxy groups 
studied by Sharina et al. has been added to the sample.  As a result
the present discussion deals with data that are homogeneous
in the sense that they refer to objects in similar environments.
The  main conclusions about the luminosity distribution of
globular clusters in dwarf galaxies would have remained
unchanged (albeit at lower statistical weight) if the Local
Group globulars (see Mackey \& van den Bergh 2005) had not 
been added to the sample.  These authors also noted that some 
of the characteristics of the globular clusters in Local Group 
dwarfs are shared by the Galactic globulars with $R_{gc} > 15$ kpc,
i.e. those that lie outside of the main body of the Galaxy. This 
observation is consistent with the view that a significant fraction 
of this cluster population component consisting of objects that 
were tidally captured from now disrupted dwarf companions  to 
the Milky Way System.

\section{LUMINOSITY DISTRIBUTION OF GLOBULARS}

 Data on the luminosity distribution on globular clusters 
in various environments are summarized in Table 2 and are 
plotted in Figures 1 - 4. Figure 1 shows the normalized 
distribution of the luminosities of globular clusters 
associated with dwarf host galaxies having $M_{v} > -16$ that are 
situated at distances D $<$ 4.0 kpc. For such objects the sample 
should be reasonably complete down to $M_{v} \sim -5.0$. The figure 
shows a luminosity distribution that peaks near the 
completeness limit and then decreases monotonically towards 
higher luminosities. A Kolmogorov-Smirnov test shows that there 
is only a 0.6\% probability that the sample of globulars in dwarf
galaxies at D $<$ 4.0 Mpc, and the sample of Galactic globulars 
with $R_{gc} < 15$ kpc (which is plotted in Figure 4), were drawn from 
the same parent population of globular cluster luminosities. 

Could the high frequency of faint globulars in dwarfs be due to
the inclusion of highly reddened young open clusters in the sample? 
This appears very unlikely because all of the host galaxies for 
objects plotted in the figure are faint galaxies with $M_{v} > -16.0$. 
Such dwarfs are all probably quite metal-poor and are therefore 
expected to contain little dust. In other words the clusters with 
$(V - I)_{o} > 0.7$  are unlikely to be reddened young open clusters. 
The most luminous object in the sample (shown as an arrow) is 
NGC 6715 = M54, which is thought (Ibata, Gilmore \& Irwin 
1994) to be the remnant nucleus of the Sagittarius dwarf. 

Figure 2 shows the luminosity distribution for the globular 
clusters associated with the more distant host galaxies that lie at
D $\geq$ 4.0 Mpc.  Probably due to incompleteness at faint magnitudes, 
the luminosity distribution for these objects peaks at a brighter 
value ($M_{v} \sim -6.2$) and then declines towards higher luminosities. 
The brightest object in this sample (which is plotted as an arrow) 
appears to be the nucleus of KK84, which has $M_{v} = -9.68$. 

Figure 3 shows the normalized luminosity distribution for the 
globular clusters in the LMC ($M_{v}= -18.5$) and the SMC ($M_{v} = -17.1$). 
The globulars associated with these objects appears to 
have a peak at $M_{v} = -7.0$ or $M_{v} = -7.5$, i.e. the globular clusters 
in the Magellanic Clouds have a similar luminosity distribution to 
that of Galactic globular clusters with $R_{gc} < 15$ kpc, the luminosity 
distribution of which is shown in Figure 4. The luminosity distribution 
of the globular clusters in the main body of the Milky Way resembles  
that of the globulars  in (mostly luminous) distant host galaxies  for 
which Harris (1991) found a peak frequency at $M_{v} \sim -7.5$. A Kolmogorov-Smirnov test shows no statistically significant difference between the luminosity distribution of the 17 globular clusters in the Magellanic 
Clouds and the luminosity distribution of the 110 Galactic globulars with 
$R_{gc} < 15$ kpc. The most luminous object in Fig. 4, which is plotted 
as an arrow, is $\omega$ Centauri, which is believed (e.g. Bekki \& 
Freeman 2003; Tsuchiya, Dinescu \& Korchagin 2003, and references 
therein) to be the nucleus of a now defunct dwarf galaxy. 

The data discussed above  show a strong dependence of the 
luminosity distribution of  globular clusters on the absolute 
magnitudes of their parent galaxies. Giant and supergiant 
galaxies, i.e hosts with $M_{v} < -16$, appear to also have luminous 
globulars with a distribution that peaks near $M_{v} \sim -7.5$. On the 
other hand faint host galaxies with $M_{v} > -16$ exhibit a luminosity 
distribution that  increases monotonically towards the 
observational completeness limit at $M_{v} \sim -5$. It is not yet clear 
to what extent the observed difference between the luminosity 
distributions of globular clusters in luminous and dim galaxies 
are intrinsic, or shaped by environmental factors. In other 
words are these differences due to the different galaxy formation 
scenarios (hierarchical mergers versus gas inflow), or were they 
the result of environmental factors, such as the preferential 
destruction of low mass clusters by the disk/bulge shocks 
(e.g. Aguilar, Hut \& Ostriker 1988) in massive giant luminous 
galaxies?

The luminosity distribution of the globular clusters in faint 
hosts exhibits some similarities to that of the globular clusters 
in the outer halo of the Galaxies (van den Bergh 2000, p.229,
Mackey \& van den Bergh 2005). This observation lends  support to 
the view that the globular cluster population in the outer Galactic 
halo at $R_{gc} > 15$ kpc consist of a mixture of (1) globular clusters 
that were initially associated with the main body of the Galaxy (2) 
other clusters that originally formed in galaxies with $M_{v} > -15$ and 
(3) a small admixture of very luminous objects that are the remnant 
cores of now defunct dwarf galaxies.

The vast majority (27 out of 30) of the blue  (presumably open)
star clusters with  $(V - I)_{o} < 0.7$,  that are listed by Sharina et al. 
(2005), are situated in  only two of the 57 objects that these authors
studied. They are the dwarf  irregular galaxies Holmberg IX
($M_{v}$ = -13,8, D = 3.7 Mpc) and UGC 3755 ($M_{v} = -15.36$, D = 5.2 Mpc).
Most of these blue clusters have absolute magnitudes in the range
$-5.5 > M_{v} > -8.0$, where the lower luminosity limit is  set by
observational selection effects.

\section{DEPENDENCE OF LUMINOSITY ON PARENT GALAXY TYPE}

 Almost all of the parent galaxies listed in Table 1 are
either late-type objects with de Vaucouleurs types
T = 6 to T = 10 (Scd - Ir) or early-type galaxies with T = -5 to T = -2
(E - S0). It is of interest to enquire if the luminosity distribution
of globular clusters depends on their parent galaxy type. Such a
dependence might perhaps have been expected if the early evolutionary 
history of galaxies is strongly correlated with their morphological
types. Unexpectedly,  examination of the data  in Table 1  shows no 
obvious differences between the luminosity distributions of globular
clusters in early-type and in late-type galaxies. However, an 
important caveat is that the present data sample is small.
The very strong dependence of the globular cluster luminosity 
distribution on parent galaxy luminosity might therefore mask
a weak dependence of the cluster luminosity distribution on host galaxy
morphological type.

\section{THE LUMINOSITY HALF-LIGHT RADIUS RELATION}

In previous papers (van den Bergh \& Mackey 2004, Mackey \& van den Bergh 2005) we have shown that few Galactic globular clusters lie above and to the left of the line 

\hspace*{5cm} Log $R_{h} = 0.25 M_{v} + 2.95$. \hspace{5cm}(1)
   
The clusters that do fall above the relation defined by Eqn. (1) 
are mostly objects suspected of being the cores of now defunct 
dwarf galaxies. Do these same conclusions also apply to the 
globular clusters associated with lower luminosity hosts? In an 
attempt to answer this question Figure 5 show a plot of the $M_{v}$ 
versus $R_{h}$ relation for globular clusters in the more-or-less 
complete sample of host galaxies with D $<$ 4.0 Mpc. The vast 
majority of the objects plotted in this figure are seen to fall 
below and to the right of the line defined by Eqn. (1). Among the 
three exceptions is NGC 6715 (=M54), plotted as a plus sign, that
is thought to be the nucleus of the Sagittarius system (Ibata, Gilmore, 
 \& Irwin, 1994), which - as expected - lies above this line.

Figure 5 also shows that the clusters associated with the LMC 
and the SMC (plotted as triangles) are systematically 
more luminous than are the globular clusters in the  
 sample of globulars associated with the nearest faint 
dwarf galaxies. Finally Figure 6 shows the following: (1) There is 
little evidence for a correlation between the half-light radii and 
the luminosities of globulars in dwarf galaxies at D $\geq$ 4.0 Mpc. 
(2) The most luminous host galaxy in our sample is UGC 3755 ($M_{v}$ = 
-15.4). The globular clusters in this galaxy (which are plotted as 
triangles) appear to have systematically smaller radii than do 
those of the globular clusters hosted by less luminous galaxies 
(which are plotted as circles). A K-S test shows only a 3\% probability 
that the data for the UGC 3755 globulars, and those associated with 
the other dwarfs with D $\geq$ 4.0 Mpc, were drawn from the same parent 
population. Finally (3) it is puzzling that the globulars in the 
dwarf sample with D $\geq$ 4.0 Mpc do {\it not} appear to follow the upper 
limit defined by Eqn. (1), whereas the globular clusters associated 
with nearer galaxies having D $<$ 4.0 Mpc (see Figure 5) mostly lie 
below this line. Contributing factors to the apparent difference
between clusters with D $<$ 4.0 Mpc and those with D$\geq$ 4.0 Mpc
might be (i) distance errors for the most distant clusters and (ii) 
systematic errors of the (small!) angular diameters measured for the
objects in the most distant cluster sample. (4) The apparent nucleus of 
KK84, which is plotted as a plus sign in Figure 5 , lies well above and to 
the left of the line defined by Eqn. (1). This suggests that this object is
indeed  the nucleus of its host galaxy. Finally  (5) it is of interest to note 
that many of the objects plotted in  Figure 6  occupy a region in the
Mv  versus Rh diagram that appears to be similar to that occupied 
 by the ``faint fuzzies'' (Larsen \& Brodie 2000) in lenticular 
galaxies. A caveat noted previously is that both the distances, and the 
cluster radius measurements are particularly uncertain for the
remote clusters having D $>$ 4.0 kpc,  It would be important to obtain 
both spectroscopic observations, and higher resolution images,  of the 
extended objects plotted in Fig.6.  Such measurements would enable one
to establish if these objects are both physically and structurally similar to
the faint fuzzies that  Larsen \& Brodie found to be associated with some 
lenticular galaxies.

\section{SPECIFIC GLOBULAR CLUSTER FREQUENCY}

The low-luminosity galaxies listed in Table 1 are remarkably 
rich in globular clusters. For the objects with D $<$ 4.0 Mpc, that 
are fainter than $M_{v} = -15.0$, one finds 38 clusters in a parent 
galaxy population that has a total luminosity $M_{v} = -15.85$. This 
yields a specific frequency (Harris \& van den Bergh 1981) of 
S = 17. Such a high value is, however, an overestimate because 
those galaxies that contain no globular clusters at all are not 
included in the total sample luminosity. However, an unbiased 
estimate of S can be obtained from a subset of the data by 
deriving the specific globular cluster frequency from the 
essentially complete Local Group data for galaxies with 
$M_{v} > -15.0$. For these objects one finds that all known 
individual Local Group (van den Bergh 2000, p.280) galaxies 
fainter than $M_{v} = -15.0$ have a combined luminosity of $M_{v} = 
-15.58$. This population contains 12 globular clusters , which 
yields a still rather high mean specific globular cluster 
frequency S = 8 $\pm$ 3. This compares to 1 $<$ S $<$ 5 for
the majority of nearby non-dwarf galaxies (Harris 1991).
Part of this difference is, no doubt, due to gas loss from
low-mass galaxies. In other words low-mass dwarf 
galaxies might actually be star poor rather than cluster rich.

 It is of interest to note that only nine of the 42 clusters  in 
Table 1 are brighter than $M_{v} = -7.5$. The high specific frequency
found for the dwarfs with D $<$ 4.0 Mpc is therefore almost entirely
due to the numerous intrinsically faint globular clusters with $M_{v} > -7.5$. 
 The results obtained above urgently raise the question whether the
high frequency of faint globulars in low-luminosity galaxies is 
intrinsic, or if it is it due to environmental factors. Was the 
formation rate of low-mass clusters particularly high in dwarfs
because they accumulated gas in a quiescent fashion? This appears 
unlikely because the quiescent dwarf irregular IC 1613 ($M_{v} = -15.3$) 
appears to contain few, if any, star clusters (Baade 1963, 
van den Bergh 1979). Alternatively one might suppose that either 
(1) the more violent processes associated with hierarchical 
merging in massive galaxies favored the formation of luminous 
globular clusters,  and/or  (2) that  the destruction of faint globulars 
was favored in massive galaxies. Perhaps the most reasonable 
interpretation of the present results is that normal cluster formed
in dwarf hosts by gradual increase in the density and pressure of the ISM, 
and/or by the slow loss of the internal energy in molecular clouds. On the 
other hand clusters in  the (starbursting) giant galaxies that generated
massive globular clusters  long ago might have formed when shocks were 
driven inward by ionization fronts generated by cosmic reionization at
z $\sim$ 6 (van den Bergh   2001, Keto, Ho \& Lo  2005), or more recently 
during violent starbursts triggered by mergers.

\section{CLUSTER LUMINOSITY SPECTRUM}
 
  If globular clusters form from the cores of giant molecular 
clouds (Harris \& Pudritz 1994, McLaughlin \& Pudritz 1996), then 
one might expect them to produce globular clusters having a power 
law luminosity distribution of the form.

\hspace*{5cm} $N(L) ~~\alpha~~ L^{\Gamma}$, \hspace*{8cm}(2) with $\Gamma$ in  the range -1.7 to -2.0.  This expectation is 
confirmed (e. g. McLaughlin 2003) for the bright end of the 
luminosity distributions of the globular cluster systems associated 
with giant galaxies. The frequency of less luminous, and hence 
less massive, clusters falls below the predictions of Equation (2). 
The reason for this  (Aguilar, Hut \& Ostriker 1988) is thought 
to be that the least massive clusters in giant galaxies are 
efficiently destroyed by the disk and bulge shocks in massive 
galaxies. The disruptive effects of disk and bulge shocks 
will be greatly reduced in dwarf galaxies in which cluster 
velocities are small and the bulge/disk masses are low. Such 
dwarfs might therefore be expected to contain globular cluster 
systems that show a power law cluster luminosity distribution 
down to quite faint cluster luminosity limits.

Figure 7 shows a plot of the distribution of the 
luminosities of the 38 globular clusters with D $<$ 4.0 Mpc, that 
are hosted by dwarf galaxies with $M_{v} > -16.0$. These clusters do 
indeed appear to exhibit a power law luminosity distribution, 
however, one that has a much shallower slope than that which is 
observed among the globulars hosted by luminous galaxies. The 
data plotted in Figure 7 unexpectedly show a power law luminosity 
distribution with $\Gamma$ $\alpha$ -0.7, rather than one with the 
expected value  $-1.7 > \Gamma > -2.0$. It is not yet clear why 
the globular clusters in faint galaxies have such a shallow 
power law luminosity distribution. 

\section{CONCLUSIONS}

Presently available data a show dramatic difference between 
the luminosity distributions of globular clusters associated with 
dwarf galaxies having parent galaxies fainter than $M_{v} = -16$, and 
the luminosity distributions of globulars hosted by luminous giant 
galaxies that are brighter than  $M_{v} = -16$. Perhaps surprisingly the 
globular clusters associated with the Magellanic Clouds (LMC  $M_{v} = 
-18.5$, SMC $M_{v} = -17.1$) resemble those hosted by giant galaxies. It 
would be particularly interesting to extend observational data on 
the globulars in UGC 3755 ($M_{v} = -15.36$) to fainter limits to see 
if this cluster-rich intermediate luminosity galaxy has a globular 
cluster system with characteristics that are intermediate between 
those surrounding giant and dwarf galaxies. 

The cluster systems hosted by giant galaxies are found to have a peak 
frequency at $M_{v} \sim -7.5$, whereas the luminosity distribution of 
globular clusters hosted by dwarf galaxies is seen to rise monotonically 
down the completeness limit of the data at $M_{v} \sim -5$. The globular 
clusters hosted by dwarf galaxies are found to have a luminosity 
distribution that may be described by a power law. However, for 
unknown reasons, this power law in dwarf galaxies is found to be 
much shallower than that which is observed among the brightest 
(most massive) globulars that are associated with giant galaxies. 

In both giant and dwarf galaxies the cores of defunct dwarf 
galaxies, such as $\omega$ Centauri and NGC 6715, lie above and to
the left of the relation given by Equation (1). Most of the clusters
hosted by the well-observed clusters hosted by galaxies at D $<$ 4 Mpc
are also found to fall below the line defined by Equation (1). It would be
interesting to obtain additional observations of the globulars in
dwarf galaxies that appear to fall in the same region of $M_{v}$ versus
$R_{h}$ space as do the ``faint fuzzies'' of Larsen \& Brodie. The 
specific globular cluster frequency in dwarf galaxies is found 
to be much higher than that in giants. This difference is almost 
entirely due to an excess of very faint globular clusters in 
dwarf Galaxies with $M_{v} > -16$. 

Unexpectedly, the present data show no evidence for a
dependence of the luminosity distribution of globular clusters
on the morphological type of their host galaxies. It will be a
challenge to theory to explain why the luminosities of globular
clusters depend so critically on the luminosity (mass?) of their
parent galaxy, while being insensitive to the morphological
type of their host galaxy.

This paper represents an extension of previous work 
done in collaboration with Dougal Mackey. I thank him for his 
wise comments and voluminous e-mail correspondence on various 
issues related to globular clusters. It is also a pleasure to 
thank Peter Stetson for writing a few special computer 
programs that were used for some calculations in the present
investigation. Finally I also wish to thank a particularly helpful
anonymous referee.

  

          
\begin{deluxetable}{llllcc}
\tablewidth{0pt}        
\tablecaption{Globular Clusters in Dwarf Galaxies} 

\tablehead{\colhead{Galaxy} & \colhead{$M_{v}$ (gal)}  & \colhead{Cluster}  & \colhead{$M_{v}$}  & \colhead{$R_{h}$(pc)}  & \colhead{D (kpc)}}

\startdata

DDO 53* T=10    &    -13.74    & 3-1120    &   -5.88    &  6.7   &  0.3\\
BKN3N* T=10     &    -9.53     & 2-863     &  -5.23     & 6.8    &  1.3\\
KDG73* T=10   &    -11.31    &  2-378    &   -5.75    & 8.3    &  1.5\\
KK77* T=-3     &    -12.21    & 4-939     &    -5.01   &  3.7   &  1.2\\
           &              & 4-1162    &   -5.37    &  6.5   &  1.6\\
           &              & 4-1165    &   -5.69    &  7.8   &  1.6\\
KDG61* T=-1    &    -13.58    &  3-1325   &   -7.55    &  4.7   &  0.0\\
KDG63* T=-3    &    -12.82    &  3-1168   &   -7.09    &  6.0   &   0.2\\
DDO78* T=-3    &    -12.75    &  1-167    &    -7.23   &  7.4   &  1.5\\

KK84 T=-3    &    14.40    &  2-785    &    -7.30    &  11.6   &  2.5\\
             &    3-705    &  -7.45    &    9.2      &  1.7    &  2.5\\
             &             & 3-705    &  -7.45       &    9.2  &  1.7 \\
             &             & 3-830    &  -9.68       &    10.6  &  0.0 \\
             &             & 3-917    &  -7.52       &    10.4  &  1.3 \\
             &             & 4-666    &  -8.37       &    10.6  &  2.4 \\
             &             & 4-967    &  -6.81       &    12.0  &  4.1 \\
             
KK112 T=10   &    -12.28   & 3-976    &  -5.93       &    9.1  &  0.3 \\
             &             & 4-742    &  -6.21       &    11.8  &  1.5 \\
             &             & 4-792    &  -6.77       &    15.0  &  1.4 \\
             
E490-017 T=10 &   -14.91   & 3-1861    &  -7.38       &    6.4  &  0.2 \\
              &            & 3-2509    &  -5.69       &    8.4  &  0.8 \\

KK065 T=10    & -13.32     & 3-1095    & -6.75       &    11.5  &   0.3\\

UGC4115 T=10  & -14.13    & 2-1042    & -6.00       &    11.9  &   1.7\\
              &           & 3-784    &  -7.53      &    9.4  &  1.2 \\
              &           & 4-1477   &   5.37      &  8.0     &  2.6\\

UA438* T=10   & -11.94    & 3-2004    & -8.67    & 3.7       &  0.2 \\
              &           & 3-3325   &  -5.96    & 3.7     &  0.4\\

UGC3755 T=10  & -15.36    & 2-652    & -8.22    & 5.5   & 1.8 \\
              &           & 2-675    &  -5.75   & 8.1   & 1.2 \\
              &           & 2-863    &  -6.93   &  5.2  &  1.0 \\
              &           & 3-739    &  -8.67   &  5.7  &  0.8 \\
              &           & 3-768    & -5.42    &  5.7  & 0.8 \\
              &           & 3-1256   & -7.77    & 8.6   & 0.4 \\
              &           & 3-1257   & -8.71    & 6.2   & 0.4 \\
              &           & 3-1611   & -7.48    & 7.5   &  0.3 \\
              &           & 3-1732   & -6.09    & 10.0  &  0.6 \\
              &           & 3-1737   &  -6.71   & 8.3   &  0.1 \\
             &            &  3-2168  &  -7.63   & 8.7   & 0.4 \\
              &          & 3-2204   &  -6.07     & 6.6  &   0.4 \\
              &          & 3-2398   &  -6.17    & 8.5   & 0.8 \\
             &          & 3-2401    &  -7.54    & 12.0   &  0.7 \\
             &          & 3-2403   &   -6.79     & 6.8    &   0.9 \\
              &         & 4-566      & -6.25     & 8.6    & 1.3 \\

LMC* T=9    & - 18.5  & NGC 1466   & -7.26    & 4.8   & 7.3 \\
            &         & NGC 1754   & -7.09    & 3.2   & 2.3 \\

BK6N* T=-3    &    -11.92    &  2-524    &    -5.40   &  4.4   &   0.8\\
           &              &  4-789    &   -5.60    &  4.5   &   1.2\\
Garland* T=10   &              &           &            &        &      \\
           &     \nodata      &  1-728    &   -8.26    &  3.2   &   1.2\\

Holmberg IX* T=10 &            & 3-1565    &   -5.31    &  4.1   &   0.1\\
             &   -13.8    &  3-1932   &   -6.61    &  9.4   &   0.4\\
             &            & 3-2373    &   -6.04    &  7.9   &   0.6\\
E540-030* T=-1    &   -11.84   & 4-1183    &    -5.37   &   6.2  &   1.6\\
E294-010* T=-3    &   -11.40   & 3-1104    &   -5.32    &  6.7   &   0.1\\
KK027*    T=-3    &   -12.32   & 4-721     &    -6.36   & 7.5    &   0.6\\
Sc22      T=-3   &   -11.10   & 2-879     &     -6.11  &  12.2  &   0.9\\
            &            & 2-100     &    -5.90   &  8.3   &    2.2\\
            &            &    4-106  &     -6.05  &  4.9   &   1.9\\
DDO113   T-10  &   -12.67   & 2-579     &    -5.60   & 7.9    &   1.5\\
            &            &  4-690    &    -5.27   &  6.5   &   0.6\\
UGC7605  T=10    &  -13.88    & 3-1503    &    -6.44   & 12.2   &   0.4\\
KK109    T=10    &   -10.19   & 3-1200    &     -5.87  &  4.4   &   1.0\\
UGC8308  T=10    &  -12.48    & 2-1198    &    -5.52   & 7.5    &   0.9\\
            &            &  3-2040   &    -6.62   & 9.1    &   1.2\\
            &            &  4-893    &     -6.30  & 5.1    &   1.6\\
            &            &  4-971    &     -5.62  &  8.3   &   1.8\\
KK211*  T=-5     &  -12.58    & 3-917     &     -6.86  &  6.3   &   0.5\\
            &            & 3-149     &    -7.82   &  6.1    &  0.0\\
            &            & 2-608     &    -8.04   &  5.0    &  1.8\\
            &            & 2-883     &    -7.07   &  8.3    &  1.4\\
            &           & 2-966      &    -9.80   &  5.7    &  0.9\\
            &           & 3-1062     &    -6.10   &  9.1    &  0.4\\
KK200  T=9      &   -12.74  & 3-1696     &   -5.68    & 9.2     &  1.2\\
KK84   T=-3      &   -14.40  & 2-785      &   -7.30    & 11.6    &   2.5\\
            &           & 3-705      &  -7.45     & 9.2     &   1.7\\
            &           & 3-830      &  -9.68     & 10.6    &   0.0\\
            &           & 3-917      &  -7.52     & 10.4    &   1.3\\
            &           & 4-666      &   -8.37    & 10.6    &  2.4\\
            &           & 4-967     &   -6.81    & 12.0    &   4.1\\
KK112  T=10       &    -12.28 & 3-976     & -5.93      &  9.1    &   0.3\\
            &           & 4-742     &  -6.21     & 11.8    &   1.5\\
            &           & 4-792     &  -6.77     & 15.0    &   1.4\\
E490-017 T=10    &   -14.91  & 3-1861    &  -7.38     &  6.4    &   0.2\\
            &           & 3-2509    &  -5.69     &  8.4    &   0.8\\
KK065    T=10    &   -13.32  & 3-1095    &   -6.75    & 11.5    &   0.3\\
UGC4115  T=10    &   -14.12  & 2-1042    &   -6.00    & 11.9    &   1.7\\
            &           & 3-784     &   -7.53    &  9.4    &   1.2\\
            &           & 4-1477    &   -5.37    &  8.0    &   2.6\\
UA438*  T=10     &    -11.94 & 3-2004    &  -8.67    &  3.7     &  0.2\\
            &           & 3-3325    &  -5.96    &  3.7     &   0.4\\
UGC3755  T=10    &    -15.36 & 2-652     &  -8.22    &  5.5     &  1.8\\
            &           & 2-675     &  -5.75    & 8.1      &   1.2\\
            &           & 2-863     &  -6.93    & 5.2      &   1.0\\
            &           & 3-739     &  -8.67    & 5.7      &   0.8\\
            &           & 3-768     &  -5.42    & 5.7      &   0.8\\
            &           & 3-1256    &  -7.77    & 8.6      &   0.4\\
            &           & 3-1257    &  -8.71    & 6.2      &   0.4\\
            &           & 3-1611    &  -7.48    & 7.5      &   0.3\\
            &           & 3-1732    &  -6.09    &10.0      &   0.6\\
            &           & 3-1737    &  -6.71    & 8.3      &   0.1\\
            &           & 3-2168    &  -7.63    & 8.7      &   0.4\\
            &           & 3-2204    &  -6.07    & 6.6     &   0.4\\
            &           & 3-2398    &  -6.17    & 8.5     &   0.8\\
            &           & 3-2401    &  -7.54   & 12.0     &    0.7\\
            &           & 3-2403    &  -6.79   & 6.8      &   0.9\\
            &           & 4-566      &  -6.25   & 8.6     &   1.3\\
LMC*   T=9      &    -18.5  &  NGC 1466   &  -7.26  &  4.8    &   7.3\\
            &           &  NGC 1754  &  -7.09   &  3.2    &   2.3\\
            &           &  NGC 1786  &  -7.70   &  3.3    &   2.2\\
            &           &  NGC 1835  &  -8.30   &  2.4    &   1.2\\
            &           &  NGC 1841  &  -6.82   & 10.8    &  12.9\\
            &           &  NGC 1898  &  -7.49   &  8.4    &   0.5\\
            &           &  NGC 1916  &  -8.24   &  2.2    &   0.2\\
            &          &  NGC 1928  &  -6.06:   &  5.6:   &   0.3\\
            &          &  NGC 1939  &  -6.85:   &  7.6:   &   0.6\\
            &          &  NGC 2005  &  -7.40    &  2.7    &   0.8\\
            &          &  NGC 2019  &  -7.75    &  2.9    &   1.1\\
            &          &  NGC 2210  &  -7.51    &  3.5    &   3.8\\
            &          &  NGC 2257  &  -7.25    & 10.5    &   7.6\\
            &          &  Hodge 11  &  -7.45    &  8.6    &   4.1\\
            &          &  Reticulum &  -5.22    &  19.3   &   9.9\\
            &          &  ESO121-SC03  & -4.37   & 10.0   &   8.4\\
SMC*  T=-5       &    -17.1 &  NGC 121      & -7.89   & 5.4    &   2.4\\
Fornax* T=-5     &   -13.1  & No.1          & -5.32   & 11.8   &   1.6\\
            &          & No.2          &  -7.03  & 8.2    &   0.9\\
            &          & NGC 1049      &  -7.66  &  4.4   &   0.6\\
            &          & No.4          &  -6.83  & 3.5    &   0.2\\
            &          & No.5          &  -6.82  & 4.4    &   1.6\\
Sagittarius* T=-5  &         &  Pal. 2       &   -8.01  &  5.4   &   \nodata\\
             &   -13.8 &  NGC 4147     &  -6.16   &   2.4  &    \nodata\\
             &         & NGC 6715     &  -10.01    &  3.8   &  0.0\\
             &         &  Ter. 7      &   -5.05    &   6.6  &   2.1\\
             &         &  Arp 2       &   -5.29    &  15.9  &   2.4\\
             &         &  Ter. 8      &   -5.05    &   7.6  &   3.5\\
             &         &  Pal. 12     &   -4.48    &   7.1  &  \nodata
\enddata
\end{deluxetable}

\begin{deluxetable}{ccccc}
\tablewidth{0cm}

\tablecaption{Luminosity distributions of globular clusters in galaxies different environments.}

\tablehead{
\colhead{Mv} & 
\colhead{D $<$ 4.0 kpc} & 
\colhead{D $>$ 4.0 kpc} & 
\colhead{LMC + SMC \tablenotemark{b}}  &  
\colhead{Galaxy \tablenotemark{c}} \\
 & \multicolumn{2}{c}{{Parent $M_{v} > -17.0$ \tablenotemark{a}}} & & \colhead{R$_{gc} < 15$ kpc}}

\startdata
-10.25  &  1\tablenotemark{d} & 0  &   0                 &  1\tablenotemark{d}\\
-9.75   &  1                  &     1\tablenotemark{d}    &  0     &  1\\    
-9.25   &   0                 &     0                     &  0     &  7\\
-8.75   &   2                 &     2                     &  0     &  8\\
-8.25   &   2                 &     2                     &  2     &  13\\
-7.75   &   4                 &     5                     &  4     &  18\\
-7.25   &   4                 &     4                     &  6     &  19\\
-6.75   &   4                 &     7                     &  2     &  19\\
-6.25   &   4                 &    10                     &  1     &   9\\
-5.75   &   5                 &     9                     &  0     &   5\\
-5.25   &  11                 &     3                     &  1     &   2\\
-4.75   &   0                 &     0                     &  0     &   3\\
-4.25   &   1                 &     0                     &  1     &   2\\   
$>$-4.0 &   0                 &     0                     &  0     &   3\\
total   &   38                &     43                    &  17    &  110\\

\tablenotetext{a}{Data from Table 1}
\tablenotetext{b} {Data from van den Bergh \& Mackey (2004)}
\tablenotetext{c} {Data from Mackey \& van den Bergh (2005)}
\tablenotetext{d} {Galaxy nucleus?}
\enddata
\end{deluxetable}

\clearpage

\begin{figure}
\caption{Normalized frequency distribution of globular clusters with D $<$ 4.0 kpc that are located in parent galaxies with $M_{v} > -16$. The luminosity function of these clusters is seen to rise monotonically down to the completeness limit at $M_{v} = -5.0$. The object marked with an arrow is NGC 6715, which is thought to be the nucleus of the Sagittarius dwarf.}
\end{figure}

\begin{figure}
\caption{Normalized frequency distribution for globular clusters with D $\geq$ 4.0 kpc, that are situated in host galaxies that are fainter than $M_{v} -16.0$. The distribution peaks at $M_{v} \sim -6.2$. Below this value the sample is incomplete. The brightest object in the sample is  the cluster 3-830, which appears to be the nucleus of KK84.}
\end{figure}  

\begin{figure}
\caption{Normalized luminosity distribution for the globular clusters in the LMC ($M_{v} = -18.5$) and the SMC ($M_{v} = -17.1$). The sample is seen to peak at $M_{v} \sim -7.3$, i.e. close to the  peak value for the globular clusters in the inner region of the Milky Way System ($M_{v} = -20.9$:).}
\end{figure}  

\begin{figure}
\caption{ Normalized luminosity distribution for the Galactic globular clusters with $R_{gc} < 15$ kpc. These data exhibit a peak at $M_{v} \sim -7.0$. The brightest object in the plot,  which is marked with an arrow, is $\omega$ Centauri - which is thought to be the stripped nucleus of a dwarf galaxy.}
\end{figure}  

\begin{figure}
\caption{ Relation between $M_{v}$ and $R_{h}$ for globulars in faint hosts that are located at D $<$ 4.0 Mpc. The figure shows that most of the clusters in these nearby dwarfs lie below and to the right of the line defined by Eqn. (1). The cluster NGC 6715, which is believed to be the nucleus of the Sagittarius dwarf spheroidal galaxy, is plotted as a plus sign. The globulars in the Magellanic Clouds (which are plotted as triangles) are seen to be systematically more luminous than those in the less luminous nearby host galaxies.}
\end{figure}  

\begin{figure}
\caption{ Relation between $M_{v}$ and $R_{h}$ for globulars in faint hosts with D $>$ 4.0 Mpc. These data show no obvious correlation between cluster luminosity and cluster radius.  Furthermore, a significant fraction of all of these clusters lie above and to left of the line defined by Eqn. (1). In this respect these clusters differ from those in the Galaxy and at D $<$ 4.0 Mpc.  The plus sign is the nucleus of the dwarf galaxy KK84. The dotted arrows show the region containing the ``faint fuzzies'' in lenticular galaxies. Many of the objects in the plot are seen to fall within the domain occupied by the ``faint fuzzies''. The globulars hosted by UGC 3755 (which is the brightest host in this sample) are plotted as triangles. Note that these objects in UGC 3755 appear to be systematically smaller than those in the other less luminous dwarfs.}
\end{figure}  

\begin{figure}
\caption{ Luminosity distribution of 38 globular clusters associated with dwarf galaxies having $M_{v} > -16.0$ and D $<$ 4.0 Mpc. The clusters are seen to have a power law distribution down to the completeness limit at $M_{v} \sim -5.0$. The distribution has an exponent $\Gamma$  $\sim$ -0.7, which is much flatter than the values $-1.7 > \Gamma > -2.0$ that are encountered in the globular cluster systems hosted by giant galaxies. The luminosities in the figure have been normalized in such a way that an object with $M_{v} = 0.0$ has L = 1.}
\end{figure}  

\end{document}